\documentclass[10pt,letterpaper,twocolumn,english]{revtex4}
\usepackage[T1]{fontenc}
\usepackage[latin9]{inputenc}
\usepackage{float}
\usepackage{graphicx}
\usepackage{amssymb}

\makeatletter

\usepackage{sfmath}
\usepackage{color}
\usepackage{graphicx}

\usepackage{babel}
\makeatother
\begin{document}

\title{Suppressing noise-induced intensity pulsations in semiconductor lasers
by means of time-delayed feedback}

\author{Valentin Flunkert}

\author{Eckehard Sch\"oll}

\affiliation{Institut f\"ur Theoretische Physik, Technische Universit\"at Berlin,
Hardenbergstra\ss{}e 36, 10623 Berlin, Germany}

\begin{abstract}
We investigate the possibility to suppress noise-induced intensity
pulsations (relaxation oscillations) in semiconductor lasers by means
of a time-delayed feedback control scheme. This idea is first studied
in a generic normal form model, where we derive an analytic expression
for the mean amplitude of the oscillations and demonstrate that it
can be strongly modulated by varying the delay time. We then investigate
the control scheme analytically and numerically in a laser model of
Lang-Kobayashi type and show that relaxation oscillations excited
by noise can be very efficiently suppressed via feedback from a Fabry-Perot
resonator.
\end{abstract}
\maketitle

\section{Introduction}

In many dynamical systems noise plays an important role and influences
the system's properties and the dynamic behavior in a crucial way.
Control of the noise-mediated dynamic properties is a central issue in nonlinear
science \cite*{SCH07}.

An often encountered effect of noise is the excitation of irregular
stochastic oscillations under conditions
where the deterministic system would rest in a stable steady state, e.g., a stable focus. 
The random
fluctuations then push the system out of the steady state. These noise-induced
oscillations are a widespread phenomenon and appear, for instance,
in lasers \cite*{PET91,DUB99,GIA00,SHE03,USH05}, chemical reaction systems \cite*{BEA05},
semiconductor devices \nocite{STE05} \nocite{HIZ06} \cite*{HIZ06,STE05}, neurons \cite*{LIN04}, and many other systems.

In practical applications the need arises to control the oscillations, for instance, by 
increasing their coherence and thus the regularity of the oscillations. 
In recent years different methods to control stochastic systems have been developed,
and applied to noise-induced oscillations in a pendulum with a randomly vibrating 
suspension axis and external periodic forcing \cite{LAN97},
stochastic resonance \cite*{GAM99,LIN01},
noise-induced dynamics in bistable delayed systems \cite{TSI01,MAS02},
and self-oscillations in the presence of noise \cite*{GOL03}. In the context of coherence
resonance \cite*{GAN93,PIK97}, time-delayed feedback in the form originally suggested
by Pyragas to stabilize unstable states in deterministic systems \cite{PYR92,SCH07}
has been demonstrated to be a powerful tool to control 
purely noise-induced oscillations \cite{JAN03}. This method
couples the difference of the actual state $X(t)\in {\mathbb R}^n$ and of a delayed
state of the system $X(t-\tau)$ back into the system.\[
\frac{d}{dt}X(t)=f(X(t),t)\,-\, K\,[X(t)-X(t-\tau)].\]
The time delay $\tau$ and the (matrix valued) control amplitude $K$
are the control parameters which can be tuned. While previous studies have shown
that the delayed feedback method can control the main frequency and the correlation time 
$t_{cor}$ and thus the regularity of noise-induced oscillations in simple systems
\cite{JAN03,BAL04,SCH04b,POM05a,POM07,PRA07,JAN07} 
as well as in spatially extended systems \cite{HIZ05,STE05a,BAL06}, 
and deteriorate or enhance stochastic synchronization of coupled systems \cite{HAU06},
in this paper we focus on the {\em suppression} of stochastic oscillations. 
We analyse the mean amplitude (or, more generally, the
covariance) of the oscillations and show that time-delayed feedback
control can decrease the mean oscillation amplitude for appropriately chosen delay time,
and thus suppress the oscillations. 

The paper is organized as follows. In section \ref{sec:Generic-model}
we study a generic model consisting of a damped harmonic oscillator
driven by white noise and investigate the influence of delayed feedback. In this
generic system we derive an analytic expression for the mean square
oscillation amplitude in dependence on the feedback and show how the oscillations
can be suppressed. In section \ref{sec:Laser-model} we consider a
semiconductor laser, a practically relevant example, and show how
optical feedback from a Fabry-Perot resonator, which realizes the
delayed-feedback scheme, can suppress noise-induced relaxation oscillations
in the laser.

\section{Generic model\label{sec:Generic-model}}

We consider a damped harmonic oscillator (whose fixed point is a stable focus) 
subject to noise ($\xi$) and feedback control \begin{eqnarray}
\dot{z}(t) & = & (\lambda-i\omega_{0})\, z(t)+D\xi(t)\qquad(z\in\mathbb{C})\label{eq:genericmodel}\\
 &  & -K\,[z(t)-z(t-\tau)],\nonumber \end{eqnarray}
where $\lambda<0$ and $\omega_{0}$ are the damping rate and the
natural frequency of the oscillator, respectively, $D$ is the noise
amplitude, $K$ is the (scalar) feedback strength and $\tau$ is the
delay time of the control term. We consider Gaussian white noise\begin{eqnarray*}
\xi(t) & = & \xi_{1}(t)+i\xi_{2}(t),\qquad(\xi_{i}\in\mathbb{R})\\
\langle\xi_{i}\rangle & = & 0,\\
\langle\xi_{i}(t)\,\xi_{j}(t')\rangle & = & \delta_{ij}\delta(t-t').\end{eqnarray*}
In our particular system the delay term in eq.(\ref{eq:genericmodel})
does not induce any local bifurcations in the deterministic system. Thus,
the fixed point is stable for all $\tau$ and $K$. 

A similar normal form (without noise) was previously used to study the
stabilization of {\em unstable} deterministic fixed points by time-delayed feedback
\cite*{HOE05}, which is possible in the same way as stabilization of 
unstable deterministic periodic orbits \cite*{SCH07,BAB02,BEC02,FIE07}.
 
\begin{figure}[H]
\centering\includegraphics[width=1\columnwidth,height=1\columnwidth,keepaspectratio]{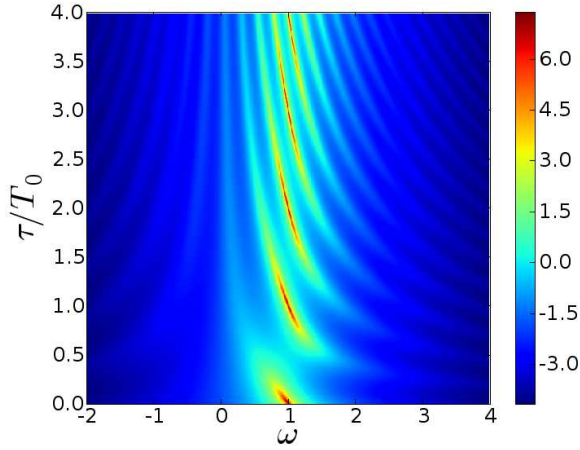}

\caption{\label{fig:fp_spec_pcolor} (color online) Power spectral density $S$ as a function of
the frequency $\omega$ and the delay time $\tau$ (logarithmic color scale). Parameters: $\lambda=-0.01$,
$\omega_{0}=1$ ($T_{0}=2\pi$), $D=1$, $K=0.2$}

\end{figure}

\begin{figure}[H]
\centering\includegraphics[width=1\columnwidth,height=1\columnwidth,keepaspectratio]{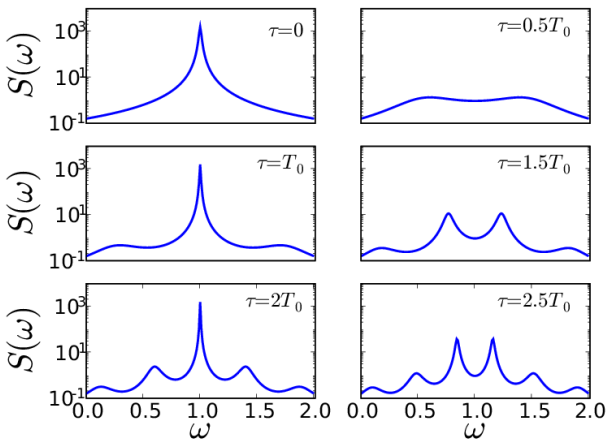}

\caption{\label{fig:fp_spec_lines} (color online) Power spectral density $S(\omega)$ for different delay
times $\tau$. Parameters: $\lambda=-0.01$, $\omega_{0}=1$, $D=1$, $K=0.2$}

\end{figure}

The power spectral density of $z$ has been calculated in \cite{POM05a}
and is given by (see Fig. \ref{fig:fp_spec_lines})\[
S(\omega)=\frac{D^{2}}{2\pi}\frac{1}{[\lambda-K(1-\cos(\omega\tau)]^{2}+[\omega-\omega_{0}+K\sin(\omega\tau)]^{2}}.\]
Figs. \ref{fig:fp_spec_pcolor} and \ref{fig:fp_spec_lines} display
the dependence of the power spectral density on the delay time. With
increasing delay new peaks appear in the spectrum. These are related
to new modes generated by the delay as will be shown later.


Before proceeding, we transform eq.(\ref{eq:genericmodel}) into a rotating
frame $z(t)=u(t)\, e^{-i\omega_{0}t}$\begin{eqnarray}
\dot{u}(t) & = & (\lambda-K)\, u(t)+K\, e^{i\omega_{0}\tau}u(t-\tau)+e^{i\omega_{0}t}D\,\xi(t)\nonumber \\
& = & a\, u(t)+b\, u(t-\tau)+D\,\tilde{\xi}(t),\label{eq:udot}\end{eqnarray}
where $a \in\mathbb{R}$, $b \in\mathbb{C}$, and $\tilde{\xi}(t)=e^{i\omega_{0}t}\xi(t)$ is a noise term with
the same properties as $\xi(t)$. The purpose of the transformation
is to make the parameter $a$ real, which will be necessary later. 

In \cite{KUE92} K\"uchler and Mensch analyzed equation (\ref{eq:udot})
for real variables. We will follow their approach and adapt it to
complex variables. Similar results for the Van der Pol oscillator have been obtained
independently in \cite{POT07}. A different two-dimensional system
with noise and delay has been recently studied in \cite{PAT06}.

We will calculate the autocorrelation function\[
G(t)=\langle u(s+t)\,\overline{u(s)}\rangle\]
in an interval $t\in[0,\tau]$, where the overbar denotes complex conjugate. 
In particular, this gives the mean
square amplitude $\langle r^{2}\rangle=\langle|z|^{2}\rangle=\langle|u|^{2}\rangle=G(0)$
of the oscillations. With the Green's function $u_{0}(t)$ solving\[
\dot{u}_{0}(t)-a\, u_{0}(t)-bu_{0}(t-\tau)=\delta(t),\]
with $u_{0}(t)=0$ for $t<0$, we can formally find a solution of
equation (\ref{eq:genericmodel})\begin{equation}
u(t)=\int\limits _{-\infty}^{t}\!\mathrm{d}t_{1}\, u_{0}(t-t_{1})\, D\,\tilde{\xi}(t_{1}).\label{eq:green_function}\end{equation}
Using (\ref{eq:green_function}) we obtain\begin{eqnarray*}
G(t) & = & \langle u(\tilde{t}+t)\,\overline{u(\tilde{t})}\rangle\\
 & = & D^{2}\int\limits _{-\infty}^{\tilde{t}+t}\! dt_{1}\,\int\limits _{-\infty}^{\tilde{t}}\! dt_{2}\, u_{0}(\tilde{t}+t-t_{1})\,\overline{u_{0}(\tilde{t}-t_{2})}\\
 &  & \times \langle\tilde{\xi}(t_{1})\,\overline{\tilde{\xi}(t_{2})}\rangle\\
 & \stackrel{s=\tilde{t}-t_{1}}{=} & 2D^{2}\,\int\limits _{0}^{\infty}\! ds\, u_{0}(s+t)\,\overline{u_{0}(s})\\
 & \equiv & 2D^{2}\, C(t).\end{eqnarray*}
The Green's function $u_{0}$ can be calculated \cite{BUD04,KUE92}
by iteratively integrating eq.~(\ref{eq:udot}) on intervals
$[k\,\tau,\,(k+1)\,\tau)$\[
u_{0}(t)=\sum_{k=0}^{\left\lfloor t/\tau\right\rfloor }\frac{b^{k}}{k!}\,(t-k\,\tau)^{k}\, e^{a\,(t-\tau k)}.\]
From the definition of $C$ and $u_{0}$ it follows that $C$ satisfies
the following equations\begin{eqnarray}
C(t) & = & \overline{C(-t)}\label{eq:Csymmetrie}\\
\dot{C}(t) & = & a\, C(t)+b\, C(t-\tau)\qquad(t>0)\label{eq:Cdot1}\\
\dot{C}(t) & = & a\, C(t)+b\,\overline{C(\tau-t)}\qquad(t>0).\label{eq:Cdot2}\end{eqnarray}
Using these three equations, we can find an ordinary differential equation for $C$,
using eqs.(\ref{eq:Cdot2}),(\ref{eq:Cdot1}),
\begin{eqnarray*}
\frac{d^{2}}{dt^{2}}\, C(t) & = & a\, C'(t)-b\,
\overline{C'(\tau-t)}\\
 & = & a\,[a\, C(t)+b\, C(t-\tau)]\\
 &  & -b\,\overline{[a\, C(\tau-t)+b\, C(-t)]}\\
 & = & a^{2}C(t)+a\, b\, C(t-\tau)\\
 &  & -a\, b\,\overline{C(t-\tau)}-|b|^{2}\overline{C(-t)}\\
 & = & (a^{2}-|b|^{2})\, C(t).\end{eqnarray*}
Here it was necessary to have a real $a$, in order for the delay
terms to cancel. Thus $C$ is of the form\[
C(t)=A\, e^{\Lambda t}+B\, e^{-\Lambda t},\]
with\[
\Lambda=\sqrt{a^{2}+|b|^{2}}=\sqrt{(\lambda-K)^{2}-K^{2}}.\]
The complex coefficients $A$ and $B$ can be found from the equations\begin{eqnarray}
C(0) & = & \overline{C(0)}\in\mathbb{R},\label{eq:boundary1}\\
\dot{C}(0) & = & a\, C(0)+b\,\overline{C(\tau)}\label{eq:boundary2}\end{eqnarray}
and\begin{eqnarray}
-1 & = & \int\limits _{0}^{\infty}\! ds\,\frac{d}{ds}\,[u_{0}(s)\,\overline{u_{0}(s)}]\label{eq:boundary3}\\
 & = & \int\limits _{0}^{\infty}\! ds\,[\dot{u}_{0}(s)\,\overline{u_{0}(s)}+u_{0}(s)\,\overline{\dot{u}_{0}(s)}]\nonumber \\
 & = & a\, C(0)+b\,\overline{C(\tau)}+a\, C(0)+\overline{b}\, C(\tau).\nonumber \end{eqnarray}
Solving equations (\ref{eq:boundary1}),(\ref{eq:boundary2}) and
(\ref{eq:boundary3}) for $A$ and $B$ gives the mean square oscillation
amplitude\begin{widetext}\begin{eqnarray}
C(0) & = & \langle r^{2}\rangle=\mathrm{Re}(A)+\mathrm{Re}(B)\nonumber \\
 & = & -\frac{1}{4\Lambda}\cdot\frac{K^{2}+2\Lambda^{2}-K^{2}\cosh(2\Lambda\tau)}{K\,\cosh(\Lambda\tau)\,[\Lambda\cos(\omega_{0}\tau)+K\,\sinh(\Lambda\tau)]+a\,[\Lambda+K\cos(\omega_{0}\tau)\,\sinh(\Lambda\tau)]}.\label{eq:r2}\end{eqnarray}
\end{widetext}\newpage

\begin{figure}[h]
\includegraphics[width=1\columnwidth]{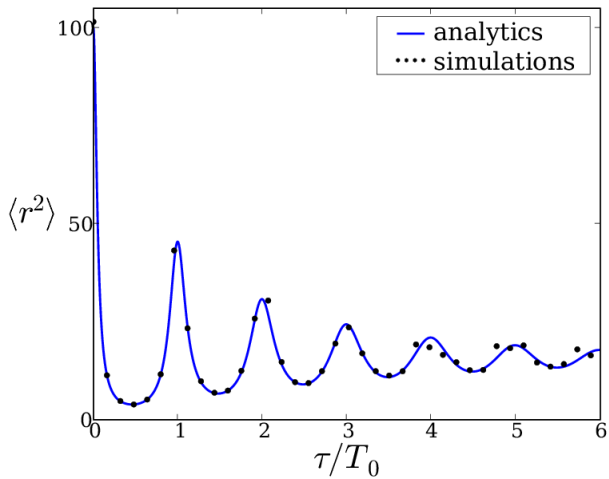}

\caption{\label{fig:r2_numerik} (color online) Mean square amplitude $\langle r^{2}\rangle$
of noise-induced oscillations (solid line: analytics, dots: numerics). 
Parameters: $\lambda=-0.01$,
$D=1$, $K=0.2$, $\omega_{0}=1$ ($T_{0}=2\pi$)}

\end{figure}

\begin{figure}[h]
\includegraphics[width=1\columnwidth]{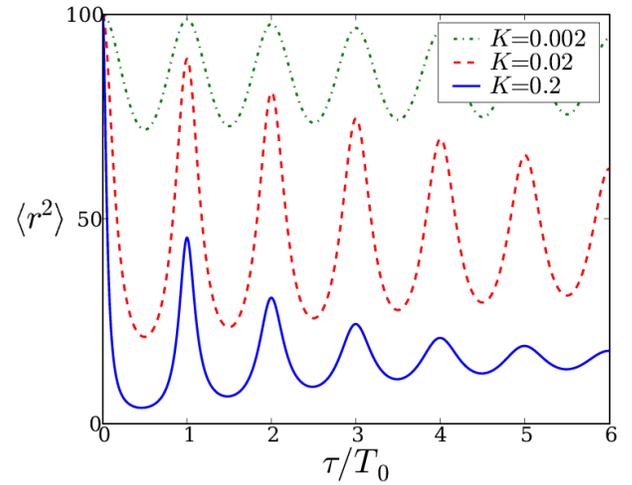}

\caption{\label{fig:r2_different_ks} (color online) Mean square amplitude $\langle r^{2}\rangle$
of oscillations as a function of the delay time $\tau$ for different
$K$ (analytic solution). Parameters: $\lambda=-0.01$, $D=1$, $\omega_{0}=1$ ($T_{0}=2\pi$)}

\end{figure}
This is now an analytic result which allows to analyze the effect
of the control term. Figure \ref{fig:r2_numerik} displays analytic
and numeric results for the oscillation amplitude. The dependence
on the control force $K$ is shown in Fig. \ref{fig:r2_different_ks}.

\begin{figure}[h]
\includegraphics[width=1\columnwidth]{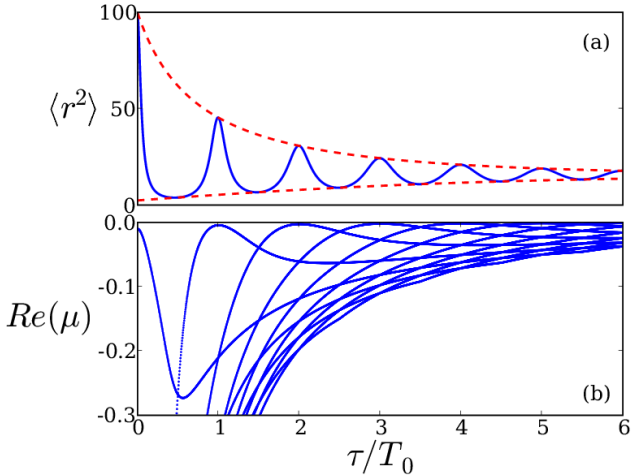}

\caption{\label{fig:r2_tau}(color online) (a) Mean square amplitude $\langle r^{2}\rangle$ of
oscillations and (b) real part of the eigenvalue spectrum of the fixed point
as a function of the delay time $\tau$. The dashed lines in (a) mark the
envelope $C_{\pm}$. Parameters: $\lambda=-0.01$,
$D=1$, $K=0.2$, $\omega_{0}=1$ ($T_{0}=2\pi$)}

\end{figure}
The oscillation amplitude can thus be strongly modulated by varying $\tau$. 
We can obtain the envelopes of the modulation by setting the terms $\cos(\omega_{0}\tau)$
to their maximum and minimum values $\pm1$ in eq.(\ref{eq:r2}) \[
C_{\pm}=\frac{1}{2\Lambda}\cdot\frac{K\sinh(\Lambda\tau)\mp\Lambda}{K\cosh(\Lambda\tau)\pm a}.\]

Figure \ref{fig:r2_tau}(a) displays $\langle r^{2}\rangle$ and the
envelopes versus $\tau$. The mean square oscillation amplitude is
modulated as a function of $\tau$ with a period $T_{0}=2\pi/\omega_{0}$. The
maxima and minima occur at\[
\tau_{+}=n\, T_{0}\quad\mbox{and}\quad\tau_{-}=\frac{2n+1}{2}\, T_{0}\]
respectively. The smallest oscillation amplitude is reached at \[
\tau_{opt}=T_{0}/2.\]

To understand the behavior of the mean square oscillation amplitude
as a function of the delay time $\tau$, one has to look at the eigenvalue
spectrum of the fixed point $z=0$ of eq. (\ref{eq:genericmodel}) (without fluctuations).
The ansatz $z(t)\propto e^{\mu t}$ in eq. (\ref{eq:genericmodel})
gives rise to a transcendental equation for the eigenvalues $\mu$:
\begin{equation}
\mu=(\lambda-i\omega_{0}-K)+K\, e^{-\mu\tau}.\label{eq:transz}\end{equation}
This equation can be solved using the Lambert function. The Lambert
function W is defined \cite{AMA05} as the inverse $W(z)$ of the equation\begin{equation}
W\, e^{W}=z\qquad(z\in\mathbb{C}).\label{eq:def_lambertW}\end{equation}
Since eq.(\ref{eq:def_lambertW}) has infinitely many solutions, the
Lambert function W has infinitely many branches $W_{n}(z)$ indexed
by $n$. Using the Lambert function W the solutions of (\ref{eq:transz})
are given by\[
\tau\mu_{n}=W_{n}[\tau\, K\, e^{-(\lambda-i\omega_{0}-K)\tau}]+(\lambda-i\omega_{0}-K)\tau.\]

Fig \ref{fig:r2_tau}(b) shows the real part of the spectrum versus
$\tau$. As $\tau$ increases, different eigenvalue branches originating from
$-\infty$ approach the zero axis (albeit remaining $<0$) and then bend away again. Since
the real part of the eigenvalues corresponds to the damping rate of
the respective mode, the oscillation amplitude excited by noise is large if a mode
is weakly damped, and small if all modes have rather large (negative) damping
rates.

\begin{figure}[H]
\centering\includegraphics[width=0.8\columnwidth,height=1\columnwidth,keepaspectratio]{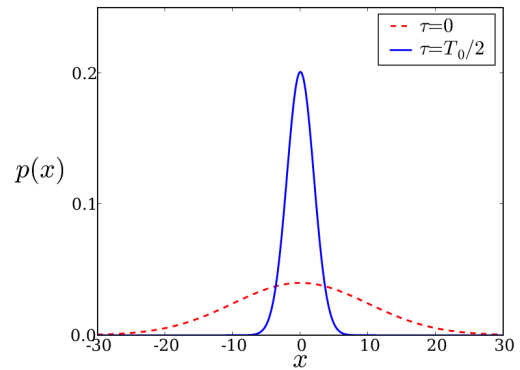}

\caption{(color online) Marginal probability distribution $p(x)$ without ($\tau=0$) and
with optimal control ($\tau=T_{0}/2$) Parameters: $\lambda=-0.01$,
$K=0.2$, $D=1$,$\omega_{0}=1$ ($T_{0}=2\pi$)\label{fig:p(x)}}

\end{figure}

Because eq. (\ref{eq:genericmodel}) is linear and $\xi$ is
Gaussian noise, the probability distribution $p(x,y)$, where $z=x+iy$,
is also a Gaussian distribution \cite*{PAT06}. The rotational invariance
($z'=z\, e^{i\phi}$) of eq. (\ref{eq:genericmodel}) implies, that
$p(x,y)$ is invariant under rotations, too. These two arguments lead
to the probability distribution\[
p(x,y)=\frac{1}{2\pi}\sqrt{\frac{1}{\sigma_{x}^{2}\,\sigma_{y}^{2}}}\,\exp\left[-\frac{x^{2}}{2\sigma_{x}^{2}}-\frac{y^{2}}{2\sigma_{y}^{2}}\right],\]
with\[
\sigma_{x}^{2}=\sigma_{y}^{2}=\frac{1}{2}\langle r^{2}\rangle.\]
Figure \ref{fig:p(x)} shows the marginal distribution 
\[
p(x)=\int\limits _{-\infty}^{\infty}\!\mathrm{d}y\, p(x,y) \nonumber
\] 
for $\tau=0$ (dashed) and $\tau=\tau_{opt}=T_{0}/2$ (solid). The case
$\tau=0$ corresponds to no control, because the feedback term in
(\ref{eq:genericmodel}) vanishes. The case $\tau=\tau_{opt}$ realizes
the optimal delay time, where the oscillations are most strongly
suppressed and the distribution $p(x,y)$ is narrowest.

\section{Laser model\label{sec:Laser-model}}

\begin{figure}[H]
\centering\includegraphics[width=0.7\columnwidth]{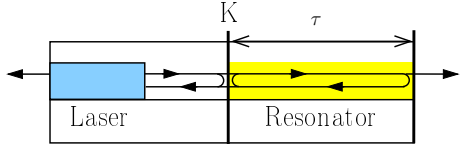}

\caption{\label{fig:laser_setup}(color online) Setup of a laser coupled to a Fabry-Perot
resonator realizing the time-delayed feedback control}

\end{figure}

In this section we investigate the effects of feedback and noise in a
semiconductor laser. A laser with feedback from a conventional mirror
can be described by the Lang-Kobayashi equations \cite{LAN80b}. Other
types of feedback have also been investigated \cite{AGR92,ERZ06}. One
particular feedback realizes the delayed feedback control with an
all-optical scheme \cite{TRO06,SCH06a}. The feedback is here generated
by a Fabry-Perot resonator. A schematic view of this setup is shown in
Fig. \ref{fig:laser_setup}. A fraction of the emitted laser light is
coupled into a resonator. The resonator then feeds an interference
signal of the actual electric field $E(t)$ and the delayed (by the
round trip time) electric field $E(t-\tau)$ back into the laser.

A modified set of non-dimensionalized \cite{ALS96} Lang-Kobayashi equations \cite{TRO06} 
describes this setup

\begin{eqnarray}
\frac{d}{dt}E & = & \frac{1}{2}(1+i\,\alpha)\, n\, E\label{eq:lk_fp}\\
 &  & -e^{i\varphi}K\,[E(t)-e^{i\psi}E(t-\tau)]+F_{E}(t),\nonumber \\
T\,\frac{d}{dt}n & = & p-n-(1+n)\,|E|^{2},\nonumber \end{eqnarray}
where the variables and parameters are defined in Table \ref{Parameters}.

\begin{table}
\begin{tabular}{c l} 
$E$ & complex field amplitude \\
$n$ & carrier density \\
$\alpha$ & linewidth enhancement factor \\
$K $ & feedback strength \\
$\tau$ & roundtrip time in the Fabry-Perot\\
$p$ & excess pump injection current\\
$T$ & timescale parameter\\
$F_E$ & noise term describing spontaneous emission\\
$\beta$ & spontaneous emission factor\\
$n_0$ & threshold carrier density\\
$\varphi, \psi$ & phases depending on the mirror positions
\end{tabular}
\caption{\label{Parameters}Nondimensionalized laser variables and parameters}
\end{table}

The phases $\varphi$ and $\psi$ depend on the sub-wavelength positioning
of the mirrors. By precise tuning $\varphi=2\pi n$ and $\psi=2\pi m$
one can realize the usual Pyragas feedback control\[
-K\,[E(t)-E(t-\tau)].\]
We consider small feedback strength $K$, so that the laser is not
destabilized and no delay-induced bifurcations occur. A sufficient condition 
\cite*{TRO06} is that\[
K<K_{c}=\frac{1}{\tau\sqrt{1+\alpha^{2}}}.\]
The noise term $F_{E}$ in (\ref{eq:lk_fp}) arises from spontaneous
emission, and we assume the noise to be white and Gaussian \[
\langle F_{E}\rangle=0,\qquad\langle F_{E}(t)\,\overline{F_{E}(t')}\rangle=R_{sp}\delta(t-t'),\]
with the spontaneous emission rate\[
R_{sp}=\beta(n+n_{0}).\]
where $\beta$ is the spontaneous emission factor, and $n_0$ is the threshold 
carrier density. Without noise the laser operates in a steady state (continuous wave 
{\em cw emission}).
To find these steady state values, we transform eqs. (\ref{eq:lk_fp})
into equations for intensity $I$ and phase $\phi$ by $E=\sqrt{I}\, e^{i\phi}$
(see Appendix A):

\begin{eqnarray}
\frac{d}{dt}I & = & n\, I-2K\,[I-\sqrt{I}\sqrt{I_{\tau}}\,\cos(\phi_{\tau}-\phi)]+R_{sp}+F_{I}(t),\nonumber \\
\frac{d}{dt}\phi & = & \frac{1}{2}\alpha\, n+K\,\frac{\sqrt{I_{\tau}}}{\sqrt{I}}\,\sin(\phi_{\tau}-\phi)+F_{\phi}(t),\label{eq:LK_I_phi_n}\\
T\,\frac{d}{dt}n & = & p-n-(1+n)\, I,\nonumber \end{eqnarray}
where $I_{\tau}=I(t-\tau)$, $\phi_{\tau}=\phi(t-\tau)$, and
\begin{eqnarray*}
\langle F_{I}\rangle=0, &  & \langle F_{\phi}\rangle=0,\\
\langle F_{I}(t)\, F_{\phi}(t')\rangle & = 0,\\
\langle F_{I}(t)\, F_{I}(t')\rangle & = & 2R_{sp}\,I\,\delta(t-t')\\
\langle F_{\phi}(t)\, F_{\phi}(t')\rangle & = & \frac{R_{sp}}{2I}\,\delta(t-t').\end{eqnarray*}

Setting $\frac{d}{dt}I=0$, $\frac{d}{dt}n=0$, $\frac{d}{dt}\phi=\mathrm{const}$,
$K=0$ and replacing the noise terms by their mean values, gives a
set of equations for the mean steady state solutions $I_{*},$ $n_{*}$
and $\phi=\omega_{*}t$ without feedback (the solitary laser mode).
Our aim is now to analyze the stability (damping rate) of the steady
state. A high stability of the steady state, corresponding to a large
damping rate, will give rise to small-amplitude noise-induced relaxation oscillations 
whereas a less stable steady state gives rise to stronger relaxation oscillations. 
Linearizing eqs. (\ref{eq:LK_I_phi_n}) around the steady state $X(t)=X_{*}+\delta X(t)$,
with $X(t)=(I,\,\phi,\, n)$ gives \begin{equation}
\frac{d}{dt}X(t)=U\, X(t)-V\,[X(t)-X(t-\tau)]+F(t),\label{eq:linearized}\end{equation}
with\begin{eqnarray*}
U & = & \left[\begin{array}{ccc}
n_{*}-\Gamma_{I} & 0 & I_{*}+\beta\\
0 & 0 & \frac{1}{2}\alpha\\
-\frac{1}{T}(1+n_{*}) & 0 & -\frac{1}{T}(1+I_{*})\end{array}\right],\\
\\V & = & \mathrm{diag}(K,\, K,\,0)\end{eqnarray*}
where $diag(...)$ denotes a $3 \times 3$ diagonal matrix, and\[
F=(F_{I},\, F_{\phi},\,0).\]
The Fourier transform of eq.(\ref{eq:linearized}) gives\[
\widehat{X}(\omega)=\underbrace{[i\omega-U+V\,(1-e^{-i\omega\tau})]^{-1}}_{\equiv M}\,\widehat{F}(\omega).\]
The Fourier transformed covariance matrix of the noise is\[
\langle\widehat{F}(\omega)\,\widehat{F}(\omega')^{\dagger}\rangle=\frac{1}{2\pi}\mathrm{diag}(2R_{sp}I_{*},\,\frac{R_{sp}}{2I_{*}},\,0)\,\delta(\omega-\omega'),\]
with the adjoint $\dagger$. The matrix-valued power spectral density $S(\omega)$
can then be defined through\[
S(\omega)\,\delta(\omega-\omega')=\langle\widehat{X}(\omega)\,\widehat{X}(\omega)^{\dagger}\rangle\]
and is thus given by\begin{eqnarray*}
S(\omega) & = & \mathrm{diag}\left(S_{\delta I}(\omega),\, S_{\delta\phi}(\omega),\, S_{\delta n}(\omega)\right)\\
 & = & \frac{1}{2\pi}\, M\,\mathrm{diag}(2R_{sp}I_{*},\,\frac{R_{sp}}{2I_{*}},\,0)\, M^{\dagger}.\end{eqnarray*}
The frequency power spectrum is related to the phase power spectrum 
$S_{\delta\phi}(\omega)$ by \cite{AGR93}\[
S_{\delta\dot{\phi}}(\omega)=\omega^{2}\, S_{\delta\phi}(\omega).\]
\begin{figure}[H]
\includegraphics[width=1\columnwidth]{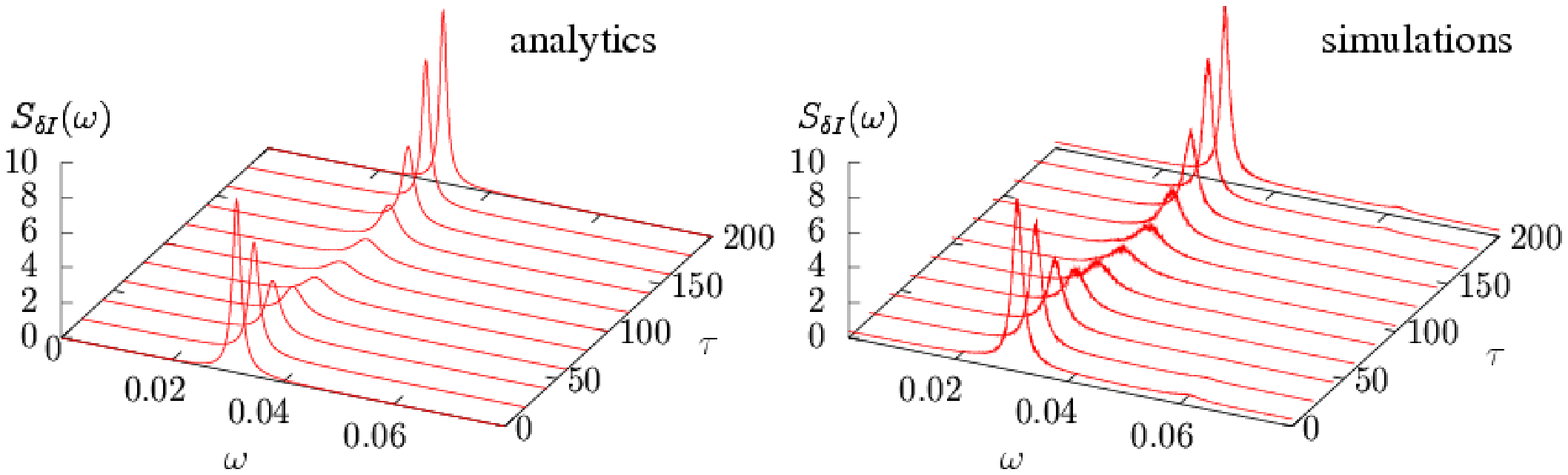}

\caption{\label{fig:specI}(color online) Analytical (left) and numerical (right) 
results for the power spectral density $S_{\delta I}(\omega)$
of the intensity for different values of the delay time $\tau$. \protect \\
Parameters: $p=1,\: T=1000,\:\alpha=2,\:\beta=10^{-5},\: n_{0}=10,\: K=0.002$.
(A typical unit of time is the photon lifetime $\tau_p=10^{-11}s$, corresponding to
a frequency of 100 GHz.)}

\end{figure}

\begin{figure}[H]
\includegraphics[width=1\columnwidth]{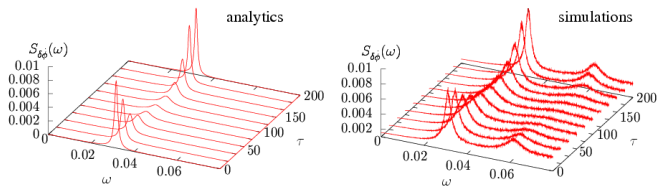}

\caption{\label{fig:specp}(color online) Analytical (left) and numerical (right) results for the power 
spectral density $S_{\delta\phi}(\omega)$ of the frequency for different values of the 
delay time $\tau$. \protect \\
Parameters: $p=1,\: T=1000,\:\alpha=2,\:\beta=10^{-5},\: n_{0}=10,\: K=0.002$}

\end{figure}
\begin{figure}[H]
\includegraphics[width=1\columnwidth]{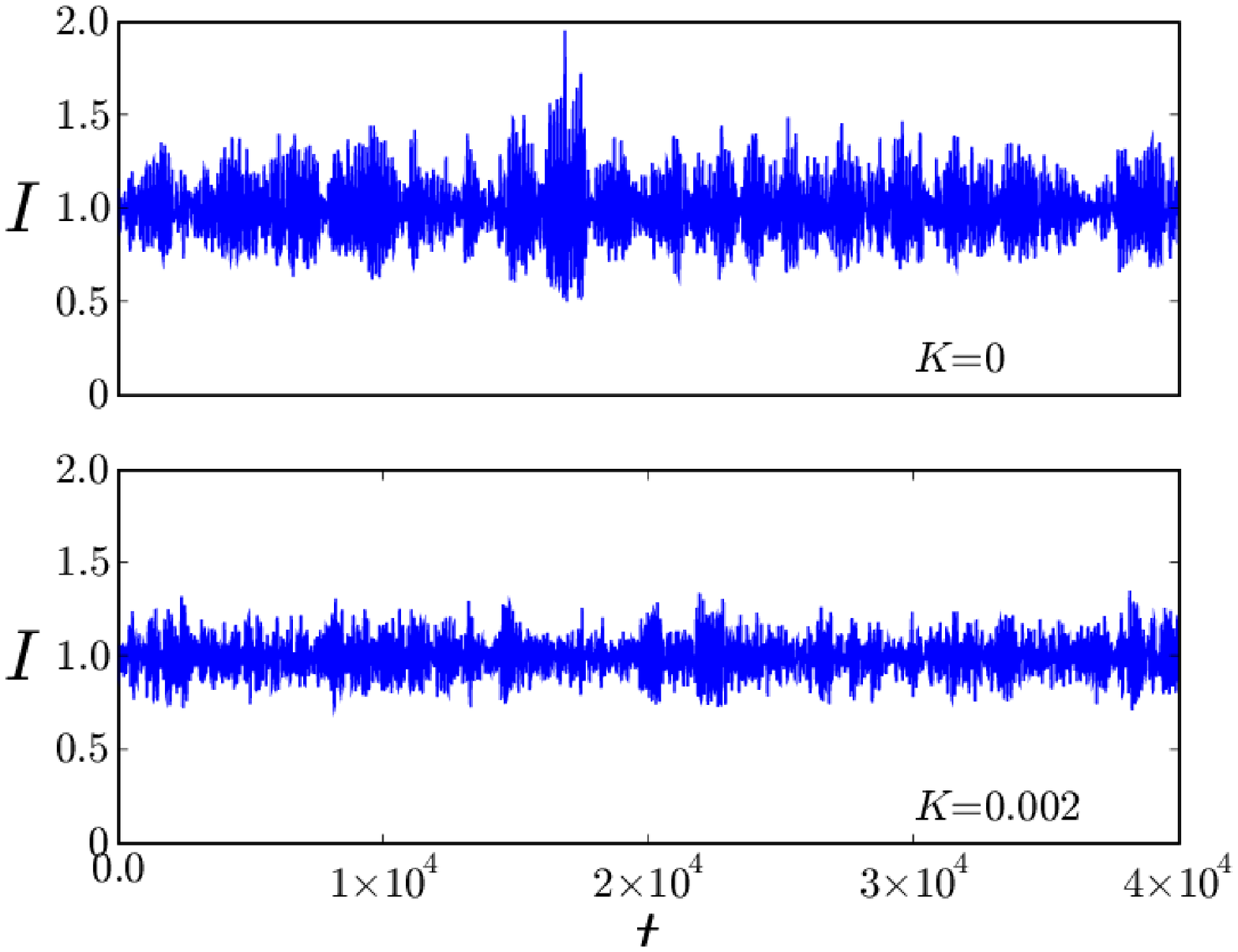}

\caption{\label{fig:lk_time_series}(color online) Intensity time series with (top panel) and without
(bottom panel) control \protect \\
Parameters: $p=1,\: T=1000,\:\alpha=2,\:\beta=10^{-5},\: n_{0}=10,\:\tau=100\approx T_{0}/2$}

\end{figure}

Figures \ref{fig:specI} and \ref{fig:specp} display the intensity
and the frequency power spectra, respectively, for different values of the delay
time $\tau$, obtained analytically from the linearized equations (left) and from
simulations of the full nonlinear equations (right). All spectra have a main peak at the 
relaxation oscillation frequency $\Omega_{RO}\approx0.03$. The higher harmonics can also
be seen in the spectra obtained from the nonlinear simulations.
The main peak decreases
with increasing $\tau$ and reaches a minimum at\[
\tau_{opt}\approx\frac{T_{RO}}{2}=\frac{2\pi}{2\Omega_{RO}}\approx100.\]
With further increasing $\tau$ the peak height increases again until it
reaches approximately its original maximum at $\tau\approx T_{RO}$.
A small peak in the power spectra indicates that the relaxation oscillations
are strongly damped. This means that the fluctuations around the steady
state values $I_{*}$ and $n_{*}$ are small. Figure \ref{fig:lk_time_series}
displays exemplary time series of the intensity with and without feedback.
The time series with feedback show much less pronounced stochastic fluctuations.

\begin{figure}[H]
\centering\includegraphics[width=1\columnwidth]{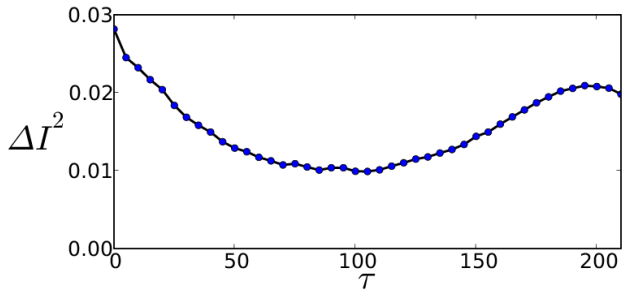}

\caption{\label{fig:Delta_I}(color online) Variance of the intensity $I$ vs. the delay time.
Parameters: $p=1,\: T=1000,\:\alpha=2,\:\beta=10^{-5},\: n_{0}=10,\: K=0.002$}

\end{figure}

Next, we study the variance of the intensity distribution as a measure for the
oscillation amplitude\[
\Delta I^2\equiv\left\langle \left(I-\langle I\rangle\right)^{2}\right\rangle .\]
This measure corresponds to the quantity $\langle r^{2}\rangle$ which we have
considered in Section II. Figure \ref{fig:Delta_I} displays
the variance as a function of the delay time. The variance is minimum
at $\tau\approx T_{RO}/2$, thus for this value of $\tau$ the intensity
is most steady and relaxation oscillations excited by noise have
a small amplitude. This resembles the behavior of the generic model
(see Fig. \ref{fig:r2_tau}(a)).

\begin{figure}[H]
\centering\includegraphics[width=0.8\columnwidth,height=1\columnwidth,keepaspectratio]{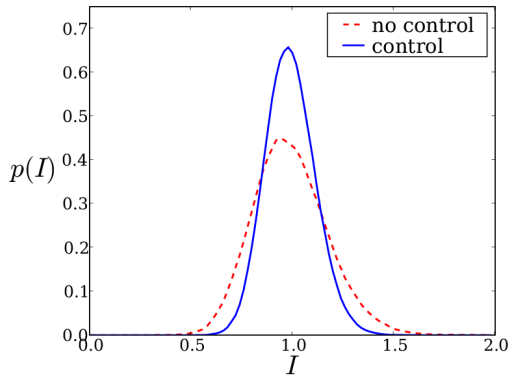}

\caption{\label{fig:lk_distr}(color online) Probability distribution of the intensity $I$
with and without the resonator (simulations). Parameters: $p=1,\: T=1000,\:\alpha=2,\:\beta=10^{-5},\: n_{0}=10,\: K=0.002$}

\end{figure}

Figure \ref{fig:lk_distr} displays the intensity distribution of
the laser without (dashed) and with (solid) optimal control (compare Fig. \ref{fig:p(x)}).
The time-delayed feedback control leads to a narrower distribution and less fluctuations.

\section{conclusion}

In this paper we have shown that time-delayed feedback can suppress 
noise-induced oscillations. 

In the first part we investigated a generic normal form model consisting of a
stable focus subject to noise and control. We found an analytic expression
for the mean square amplitude of the oscillations. This quantity is
modulated with a period of $T_{0}=2\pi/\omega_{0}$ in dependence on $\tau$.
For $\tau=T_{0}/2$ the oscillations have the smallest amplitude. 

In the second part we considered a semiconductor laser coupled to a Fabry-Perot
resonator. In the laser spontaneous emission noise excites stochastic relaxation
oscillations. By tuning the cavity round trip time to half the relaxation
oscillation period $\tau_{opt}\approx T_{RO}/2$ the oscillations
can be suppressed to a remarkable degree. This is demonstrated in the power
spectra of the intensity and the frequency, where the relaxation oscillation
peak has a minimum height at $\tau_{opt}$. The variance of the intensity distribution
$\Delta I$ shows a minimum at $\tau_{opt}$, thus the intensity distribution
is narrowest at this value of $\tau$.

\section*{Acknowledgment}

This work was supported by Deutsche Forschungsgemeinschaft in the framework of Sfb 555.
We thank Andreas Amann, Philipp H{\"o}vel, and Andrey Pototsky for fruitful discussions.

\appendix
\section{Ito transformation}
Ito's formula describes how a stochastic differential equation
(SDE) is transformed to new coordinates. Consider the stochastic
differential equation for $x(t)$\[ dx(t)=a[x(t),t]\, dt+b[x(t),t]\,
dW(t).\] Ito's formula specifies the transformation to a new variable
$y=f(x)$.  The SDE for $y$ is given by\begin{eqnarray*} dy & = &
df[x(t)]\\ & = & a[x(t),t]\, f'[x(t)]\, dt+b[x(t),t]\, f'[x(t),t]\,
dW(t)\\ &&+\frac{1}{2}\, b[x(t),t]^{2}\, f''[x(t)]\,
dW^{2}.\end{eqnarray*}

We will apply Ito's formula to rewrite the laser equations for the
complex electric field $E$ in terms of the amplitude $A$ and the phase
$\phi$ 
\[ 
E=A\, e^{i\phi}.
\] 
The equation for $E$ without feedback is given by
\[
\frac{d}{dt}E=\frac{1}{2}(1+i\alpha)\, n\, E + F_{E}(t)
\] 
or written as a stochastic differential equation
\[
dE=\frac{1}{2}(1+i\alpha)\, n\, E\,dt + \sqrt{\frac{R_{sp}}{2}}\, dW(t),
\]
with the complex Wiener process $dW=dW_{x}+i\, dW_{y}$. We define the
new coordinates 
\[ 
\mu+i\phi=\log A+\log e^{i\phi}=\log[E_{x}+i\, E_{y}].
\] 
Using Ito's formula with
\[
a[E]=\frac{1}{2}(1+i\alpha)\, n\, E,\quad b[E]=\sqrt{\frac{R_{sp}}{2}}\quad\mbox{and},\quad f[E]=\log E,
\]
we find\begin{equation}
d(\mu+i\phi)=\frac{1}{2}(1+i\alpha)\, n\,\frac{1}{E}\, dt-\frac{1}{2}\frac{R_{sp}}{2}\frac{1}{E^{2}}\, dW^{2}+\sqrt{\frac{R_{sp}}{2}}\,\frac{1}{E}\, dW.\label{eq:mu_iphi}\end{equation}
For a complex Wiener process $dW$ one can easily see that\[
dW^{2}=[dW_{x}+idW_{y}]^{2}=dW_{x}^{2}+2i\,dW_{x}dW_{y}-dW_{y}^{2}=0.\]
Here we used $dW_x ^2=dW_y ^2=dt$ and $dW_x dW_y=0$ \cite{GAR02}.
Thus, equation (\ref{eq:mu_iphi}) simplifies to 
\begin{eqnarray*}
d(\mu+i\phi)&=&\frac{1}{2}(1+i\alpha)\, n\,\exp(-\mu-i\phi)\,\, dt\\
&&+\sqrt{\frac{R_{sp}}{2}}\,\exp(-\mu-i\phi)\, dW.
\end{eqnarray*}
Splitting this equation into real and imaginary part and transforming
with Ito's formula back to $A=\exp\mu$, we obtain 
\begin{eqnarray*}
dA & = & (\frac{1}{2}\, n\, A+\frac{R_{sp}}{4A})dt+\sqrt{\frac{R_{sp}}{2}}\left(\cos\phi\, dW_{x}+\sin\phi\, dW_{y}\right)\\
d\phi & = & \frac{1}{2}\, n\,\alpha\,dt+\frac{1}{A}\sqrt{\frac{R_{sp}}{2}}\left(-\sin\phi\, dW_{x}+\cos\phi\, dW_{y}\right).
\end{eqnarray*}
Because the rotation is an orthogonal transformation, one can
understand the increments as new independent Wiener processes 
\begin{eqnarray*}
dW_{A}&=&\cos\phi\, dW_{x}+\sin\phi\, dW_{y}\\
dW_{\phi}&=&-\sin\phi\, dW_{x}+\cos\phi\, dW_{y}.
\end{eqnarray*}

We have derived the laser equations in polar coordinates. To include
the delay terms does not change the derivation and we will just state
the result here\begin{eqnarray*}
\frac{d}{dt}A & = & \frac{1}{2}nA-K[A-A_{\tau}\cos(\phi_{\tau}-\phi)]+\frac{R_{sp}}{4A}+F_{A}(t)\\
\frac{d}{dt}\phi & = & \frac{1}{2}\alpha n+K\frac{A_{\tau}}{A}\,\sin(\phi_{\tau}-\phi)+F_{\phi}(t)\end{eqnarray*}
with\begin{eqnarray*}
\langle F_{A}(t)\, F_{A}(t')\rangle & = & \frac{R_{sp}}{2}\,\delta(t-t')\\
\langle F_{\phi}(t)\, F_{\phi}(t')\rangle & = & \frac{R_{sp}}{2A^{2}}\,\delta(t-t').\end{eqnarray*}

To obtain the equations for intensity $I=f(A)=A^{2}$ instead of the
amplitude Ito's formula has to be applied again (of course this could
be done in one step from the initial equations).  The amplitude
equation is given by\begin{eqnarray*}
dA&=&\left\{\frac{1}{2}nA-K[A-A_{\tau}\cos(\phi_{\tau}-\phi)]
+\frac{R_{sp}}{4A}\right\}dt\\
&&+\sqrt{\frac{R_{sp}}{2}}\, dW(t).\end{eqnarray*} For a real
stochastic process $dW$ holds $dW^{2}=dt$. Using Ito's formula with $dW^{2}=dt$, $f'(A)=2A$ and,
$f''(A)=2$, we find
\begin{eqnarray*}
dI&=&(\frac{1}{2}nA-K[A-A_{\tau}\cos(\phi_{\tau}-\phi)]+\frac{R_{sp}}{4A})2A\,dt\\
&&+\frac{1}{4}R_{sp}2dt+\sqrt{\frac{R_{sp}}{2}}\,2A\, dW(t)
\end{eqnarray*} 
and thus
\begin{eqnarray*}
\frac{d}{dt}I&=&nI-K[I-\sqrt{I}\sqrt{I_{\tau}}\cos(\phi_{\tau}-\phi)]\\
&&+R_{sp}+F_{I}(t),\end{eqnarray*}
with\[ \langle F_{I}(t)\, F_{I}(t')\rangle=2R_{sp}I\,\delta(t-t').\]

\vfill


\end{document}